\documentclass[mathleft, fleqn]{an}
\usepackage{graphicx}
\usepackage[varg]{txfonts}
\pdfoutput=1
\begin{document}
%

\Pagespan{1}{}
\Yearpublication{2014}%
\Yearsubmission{2014}%
\Month{0}%
\Volume{999}%
\Issue{0}%
\DOI{asna.201400000}%

   \title{Excess B-modes extracted from the Planck polarization maps}

   \author{ H. U. N{\o}rgaard - Nielsen\thanks{hunn@space.dtu.dk}}

   \institute{National Space Institute (DTU Space),\\
    Technical University of Denmark, \\
    Elektrovej, DK-2800 Kgs. Lyngby, Denmark}

\titlerunning{B- modes extracted for the Planck polarization maps}
\authorrunning{H. U. N{\o}rgaard - Nielsen}
\received{XXXX}
\accepted{XXXX}
\publonline{XXXX}


 \abstract{One of the main obstacles for extracting the Cosmic Microwave Background (CMB) from mm/submm observations is the pollution from the main Galactic components: synchrotron, free-free and thermal dust emission. The feasibility of using simple neural networks to extract CMB has been demonstrated on both temperature and polarization data obtained by the WMAP satellite. The main goal of this paper is to demonstrate the feasibility of neural networks for extracting the CMB signal from the Planck polarization data with high precision. Both auto-correlation and cross-correlation power spectra within a mask covering about 63 percent of the sky have been used together with a 'high pass filter' in order to minimize the influence of the remaining systematic errors in the Planck Q and U maps.  Using the Planck 2015 released polarization maps, a  BB power spectrum have been extracted by \textit{Multilayer Perceptron} neural networks. This spectrum contains a bright feature  with signal to noise ratios $\simeq$ 4.5 within  200 $\leq$ l $\leq$ 250. The spectrum is significantly brighter than the BICEP2 2015 spectrum, with a spectral behaviour quite different from the 'canonical' models (weak lensing plus B-modes spectra with different tensor to scalar ratios). The feasibility of the neural network to remove the residual systematics from the available Planck polarization data to a high level has been demonstrated.}

\maketitle
%

\section{Introduction}
Since the discovery of the Cosmic Microwave Background radiation (CMB) by Penzias \& Wilson(1965), it has been studied intensively. It has been generally accepted that the radiation can provide unique information about the early phases of the evolution of the Universe. Observational efforts have been concentrated on improving the sensitivity and angular resolution of the CMB temperature sky maps.

It was already predicted by Rees(1968) that detection of  polarization of the CMB on large angular scales could be a very important way to investigate the very early evolution of the Universe. Furthermore, CMB polarization measurements can also give clues to the reionization phase at a much later epoch.

Fundamental symmetries in the production and growth of the polarization signal put constrains on configurations of the CMB polarization. Scalar (density) pertubations give rise to temperature (T) fluctuations and E (curl-free) polarization modes, while tensor (gravitational waves) give rise to both T fluctuations, and E and B (divergence-free) modes.

If the initial inhomogeneities in the Universe were Gaussian in nature, it follows from linear theory that the CMB fluctuations are also Gaussian and fully described by 4 power spectra  TT, EE, BB and TE, while the TB and EB power spectra vanish due to parity constraints e.g. Kamionkowski et. al. (1997). Among others, Hu \& Dodelson(2002) emphasize in their CMB review that the detection of substantial CMB B modes will be momentous and push for an introduction of new physics in CMB research.

CMB polarization measurements were provided by the WMAP satellite, the final results given in Bennett et al. (2013). For r, the tensor-to-scalar ratio, Bennett et al. found, for WMAP-only data, an upper limit of 0.38.

Ade et al. (2014) report a detection of B-modes in a 400 sq. degree area on the sky with the BICEP2 instruments, observing from the South Pole.
These observations were done at only one frequency, 150 GHz, implying that their estimate of polarized Galactic emission in this small area on the sky is uncertain. Ade et al. (2015) attacked this problem by combining BICEP2 + Keck 150 GHz data and Planck 30 GHz-353 GHz observations in this area.  This investigation determined (50 $\leq$~l$\leq$ 200) the amplitude of the lensing  spectrum 1.12, relative to the standard $\lambda$CDM model, with a detection significance of 6.8 $\sigma$  and an upper limit on the tensor-to-scalar ratio r $<$ 0.13, with 95 \%\ confidence.

The ESA Planck mission was launched in May 2009 and covered the sky nearly 5 times with unprecedented sensitivity and angular resolution. The temperature data for the first 15 months together with a series of papers  were released in March 2013. Both the LFI and HFI detector systems were sensitive to CMB temperature and polarization fluctuations. The first Planck polarization data (analysed in this paper) was released to the general astronomical community  in February 2015. The maps and auxiliary data can be found in the Planck Legacy Archive (http://www.cosmos.esa.int/web/planck/pla).

In connection with the first release of Planck temperature data, the results of testing of 4 component separation methods (COMMANDER, NILC, SEVEM and SMICA) was summarised in Planck Collaboration XII (2014). Similarly, the results of these methods obtained for the released polarization data was published simultaneously with the data release in February 2015 (Planck Collaboration IX, details of the 4 methods are given).
Concerning B-modes, the BB power spectra from these methods was cross correlated (50 $<$ l $<$ 1100) with the Planck $\lambda$CDM lensing model and amplitudes of 1.00$\pm0.10$ has been found.

The Planck Collaboration XI (2015) updates the parameter estimates given in Planck Collaboration XV (2014) . An upper limit on the tensor-to-scalar ratio of 0.11 was found, similar to the limit given by BICEP2, Ade et al. (2015)

Development of detailed simulations of the whole Planck data reduction pipeline, from raw observations including including detailed models of sources of  systematics, to all-sky frequency maps, has been an essential part of the mission. This effort has been coordinated by the Planck 'Component Separation' Working Group (led by Drs. M. Ashdown and C. Baccigalupi). Several series of simulations have been performed, including more and more complex modeling of the systematic errors in the data.
As the mission progress, the simulations have been obtained by exploiting a specially developed software package called Planck Sky Model (Delabrouille et al. 2013). The latest available full scale simulation data set, called FFP8, has been used in this paper.

The advantage of analysing simulations is that everything is known about the data, including noise characteristics and instrumental effects. Therefore, it is possible to investigate in great details the effects of the remaining systematic errors in the derived CMB maps. Furthermore, a great effort has been performed to ensure that these simulations simulate the different astrophysical components in the sky, and the known sources of systematic errors in the observations. In this investigation, the main purpose of using these simulations is to obtain accurate calibration of the power spectra derived from the extracted CMB maps.

In a series of papers the capabilities of neural networks for dealing with mm/submm observations have been investigated by N{\o}rgaard - Nielsen \& J{\o}rgensen (2008),  N{\o}rgaard - Nielsen \& Hebert (2009),
N{\o}rgaard - Nielsen (2010), and N{\o}rgaard - Nielsen (2012).
The last two papers showed, by using data from the WMAP satellite, that simple neural networks can extract the CMB temperature and polarization signals with excellent accuracy.

As emphasized in these papers, the main advantages of neural networks for extracting CMB signals from submm/mm observations are:
\begin{itemize}
\item{the small number of weights to be determined in order to set up the full transformation between the input frequency maps and the CMB output quantity, either the CMB Stokes parameter Q or U (short Q and U in the following). In the neural networks used in this paper, the number of weights is $\sim$ 80}
\item{the foreground model is only using the input frequency maps themselves (neither auxiliary data nor assumptions about e.g. the spectral behaviour of the different foreground components are applied)}

\item{a small number of parameters are needed in the foreground model, since a physical meaningful model is not required }

\item{the network is set up to extract signals with the same spectral behaviour as CMB}

\item{the available frequency range is exploited}

\end{itemize}

In the analysis of the WMAP 7yr polarization data, a faint BB power spectrum was detected. It was shown that it was not a result of pollution from the polarized foregrounds, nor introduced by the neural networks themselves. Due to the improvement in sensitivity by the neural network method, it is most probable  that the detected BB power spectrum is caused by some systematic errors not recognized in previous analyses of these polarization maps.

The paper is structured in the following way.  A brief description of the neural network concept is given in Sect. 2, together with details of the specific neural networks used in this analysis. Sect. 3 contains a description of the observed and simulated Planck data, and how the non - CMB foregrounds are modelled. Sect. 4 shows the extracted Q and U maps. Sect. 5 gives details about how the auto and cross correlated EE and BB power spectra have been derived. The calibration procedure is outlined in Sect. 6. The final power spectra are given in Sect. 7. The conclusions are given in Sect 8.

\section{The neural networks}
\subsection{Brief description of  the neural network concept}

The basic concept of the neural network method for extracting CMB signals from microwave observations has been outlined in our previous papers N{\o}rgaard - Nielsen \& J{\o}rgensen (2008),  N{\o}rgaard - Nielsen \& Hebert (2009),
N{\o}rgaard - Nielsen (2010), N{\o}rgaard - Nielsen (2012). An excellent introduction to the many different types of neural networks can be found in Bishop (1995).

The basic idea of a neural network is that it provides a transformation between a number of input channels (here spectra containing fluxes at different frequencies) and a number of output channels (here CMB Q or U fluxes).

In all our papers, one of the simplest and also most popular network has been exploited, namely the Multilayer Perceptron (MLP).

\begin{figure}[h]
\raggedright
\includegraphics[width = 3.0 in]{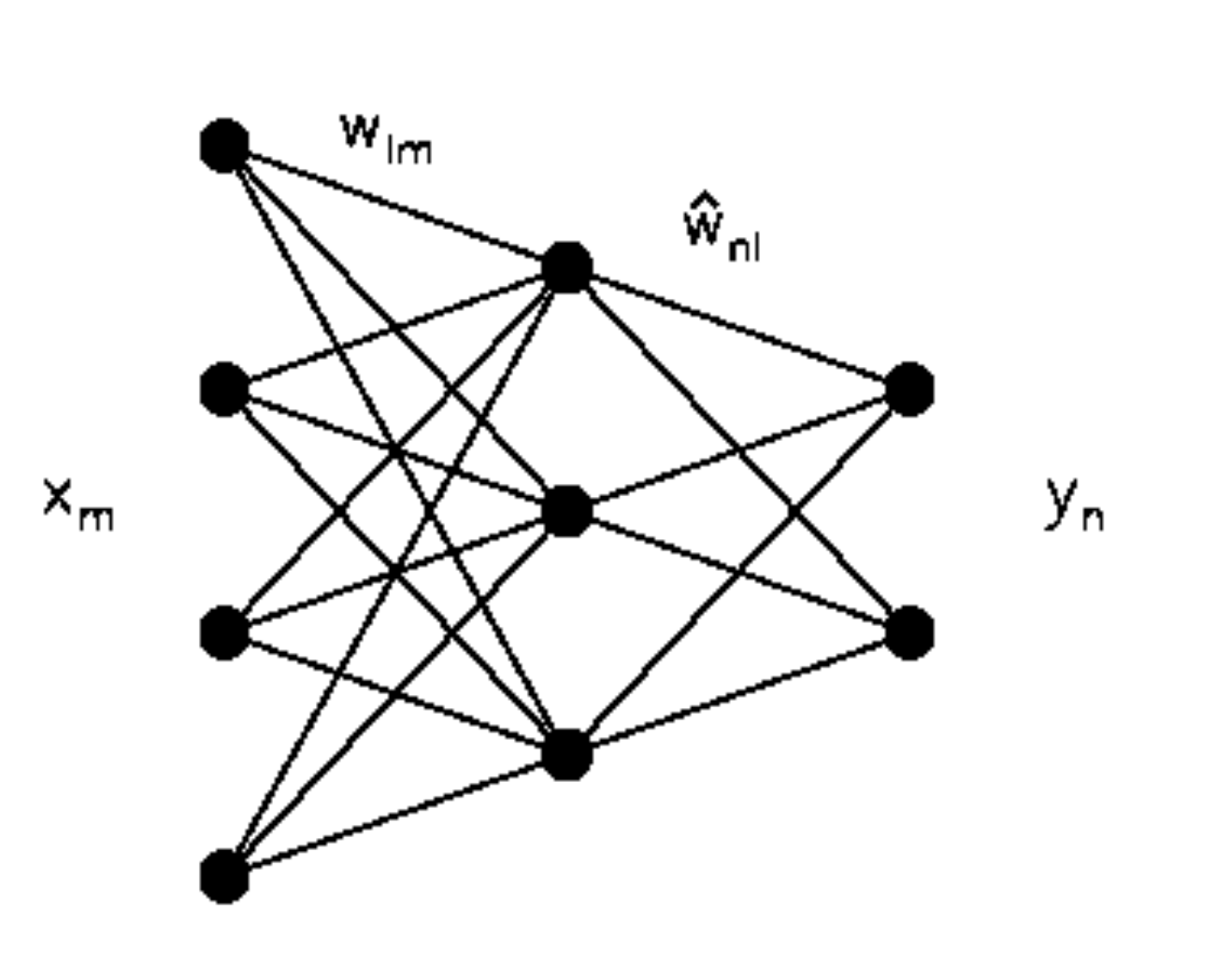}
\caption{A schematic of a Multi Layer Perceptron network with
one hidden layer.}
\label{fig_1}
\end{figure}

An MLP consists of a network of units (called processing elements, neurons or nodes) as illustrated in Fig.~\ref{fig_1}. Each unit is shown as a circle and the lines connecting them are known as weights or links. The network can be understood as an analytical mapping between a set of input variables $x_{m}~(m = 1,...,M)$  and a set of output variables $y_{n}~(n=1,...,N)$. The input variables are applied to the M input units on the left of the figure: M = 4 and N = 2 in the shown example. These variables are multiplied by a matrix of parameters $w_{lm}~(l = 1,...,L);~
m=(1,...,M)$ corresponding to the first layer of links. Here L is the number of units in the middle (hidden) layer: L = 3 in the example shown.


This results in a vector of inputs to the units in the hidden layer. Each component of this vector is then transformed by a non-linear function F, giving

\begin{equation}
    z_{l}~=~F \left( \sum_{m=1}^{M}~w_{lm}x_{m}~+\Theta_{l} \right) ~~(l=1,...,L), \label{eq1}
\end{equation}

where $\Theta_{l}$ is an offset or threshold. The Neural Network Toolbox in the MATLAB software environment ($www.mathworks.com$)  has been used with the $tansig$ function as the non-linear function:

\begin{equation}
    \mathrm{tansig}(x) ~= ~ \frac{2}{1~+~\mathrm{exp}(-2~x)}~-1, \label{eq2}
\end{equation}

It is seen that \emph{tansig} has an S-shape, with values falling within the interval $[-1:1]$.
From the hidden layer to the output units a linear transformation with weights $\widehat {w}_{nl}~(n=1,...,N;l=1,...,L)$ and offsets $\widehat{\Theta}_{n}$ are applied

\begin{equation}
y_{n}~=~\sum_{l=1}^{L}\widehat{w}_{nl}z_{l}~+~\widehat{\Theta}_{n}
\quad\quad (n=1,...,N), \label{3}
\end{equation}
By combining Eqs.\,1 and 2 it is seen that the entire network transforms the inputs $x_{m}$ to the outputs $y_{n}$ by the following analytical function
\begin{equation}
y_{n}(x_{1},...,x_{M})~=~ \sum_{l=1}^{L}\widehat{w}_{nl}~F \left( \sum_{m=1}^{M}w_{lm}x_{m}~+~\Theta_{l}\right)
~+~\widehat{\Theta}_{n}, \label{eq4}
\end{equation}

Clearly, such an MLP can easily be generalized to more than one hidden layer.

Given a set of P example input and output vector pairs $\{x_{m}^{p}~y_{n}^{p}\}~ p=1,...,P$ for a specific mapping, a technique known as error back propagation, can derive estimates of the parameters $w_{lm},~\Theta_{m}$ and   $\widehat{w}_{nl},~\widehat{\Theta}_{n}$, so that the network function (\ref{eq4}) will approximate the required mapping.

The training algorithm minimizes the error function
\begin{equation}
E_{NET} ~=~\sum_{p=1}^{P}\sum_{n=1}^{N}[y_{n}(x^{p}) ~- ~y_{n}^{p}]^{2}, \label{eq5}
\end{equation}

In order to set up a neural network a training data set is created (details are given in Section 5). Assuring that the parameters of the input spectra are covering the parameter intervals of the data set to be analysed, a total of say 50.000 spectra are calculated and appropriate noise maps are added.

This training set is then input to the network. The neural network will then find the required relation. The accuracy of the obtained network is evaluated by means of a completely independent test data set, constructed in the way as the training set but with independent noise maps.

 Due to the statistical nature of setting up the initial weights of the network, several networks are investigated each time. The main selection parameter for choosing the network to be used is that the distribution of residuals between the true output values and the derived values are Gaussian (the skewness and kurtosis of the deviations  must be $\leq$ 0.01). The rms of the residuals are also taken into account, but with a lower weight.

Once the network has the desired accuracy, the data sets to be analysed are run through the network.

\subsection{The applied neural networks}
Basic properties of the neural network method:
\begin{enumerate}
\item{the noise in the data is assumed white (e.g. no 1/f noise).}
\item{no corrections applied for the differences in angular resolutions of the input frequency maps(if the frequency maps are corrected to a common angular resolution, the extracted maps will have a significantly poorer resolution). }
\item{each sky pixel is handled totally independently of the other pixels}
\end{enumerate}

The neural networks applied here have 7 input channels, the Planck 30 GHz - 353 Ghz channels and one output channel, CMB Q or U. The 7 input values are referred to as a spectrum. The setup of a neural network follows this scheme:
\begin{enumerate}
\item{To simulate a spectrum, draw relevant number of independent parameters (in our case 2 foreground parameters (see Sect. 3) and a Q or U value), uniformly distributed within specified ranges}

\item {Calculate the resulting spectrum from the foreground model in Sect.3}

\item {For each frequency, add random Gaussian noise calculated from the Planck hit maps.}

\item {Repeat 1--3 until the desired number of spectra ($N_{NNET}$) has been obtained. This data set is split into a set used directly to train the network ($N_{TRAIN}$) and a set used for validation of the network ($N_{TEST}$).}

\item {Train the neural network to find the transformation between the input spectra and the true CMB Q or U (known for each spectrum of the training and test data sets).
}
\item {Run the $N_{TEST}$ spectra through the network to get an independent estimate of the reliability of the network, derived from the skewness and kurtosis of the distributions of residuals (e.g Q(true) - Q(NNET)) and the correlations of the residuals with the input parameters}

\item {If the derived network is working satisfactorily (the systematic errors and correlations on the independent test sample are as small as found in our previous investigations) the maps to be analysed are run through the network }

\end{enumerate}

 An MLP with 2 hidden layers (6 and 4 processing elements, respectively, referred to as the NN network) was used for each of the data sets considered here (Sect. 3).

 In such a network, only 81 weights  are needed to provide the full transformation from the input spectra to the output Q or U value. With this amount of processing elements it takes about 2 min to establish a neural net on one of the DTU Space 40 cpu computers. By changing the number of processing elements per hidden layer or the number of hidden layers is not changing the transformation significantly.
 The experience is that about 100,000 spectra are enough to for each of the training  and testing  data sets.

 As previous found in our paper on the WMAP 7 yr polarization maps,
skewnes and kurtosis of the distributions of residuals Q(true) - Q(NN) and U(true) - Q(NN) are small, $\leq$ 0.03, and the correlations of these residuals with the input parameters are also small,  $\leq$ 0.05.


\begin{figure}[h]
\centering
\includegraphics[angle = 180, width=3.0in]{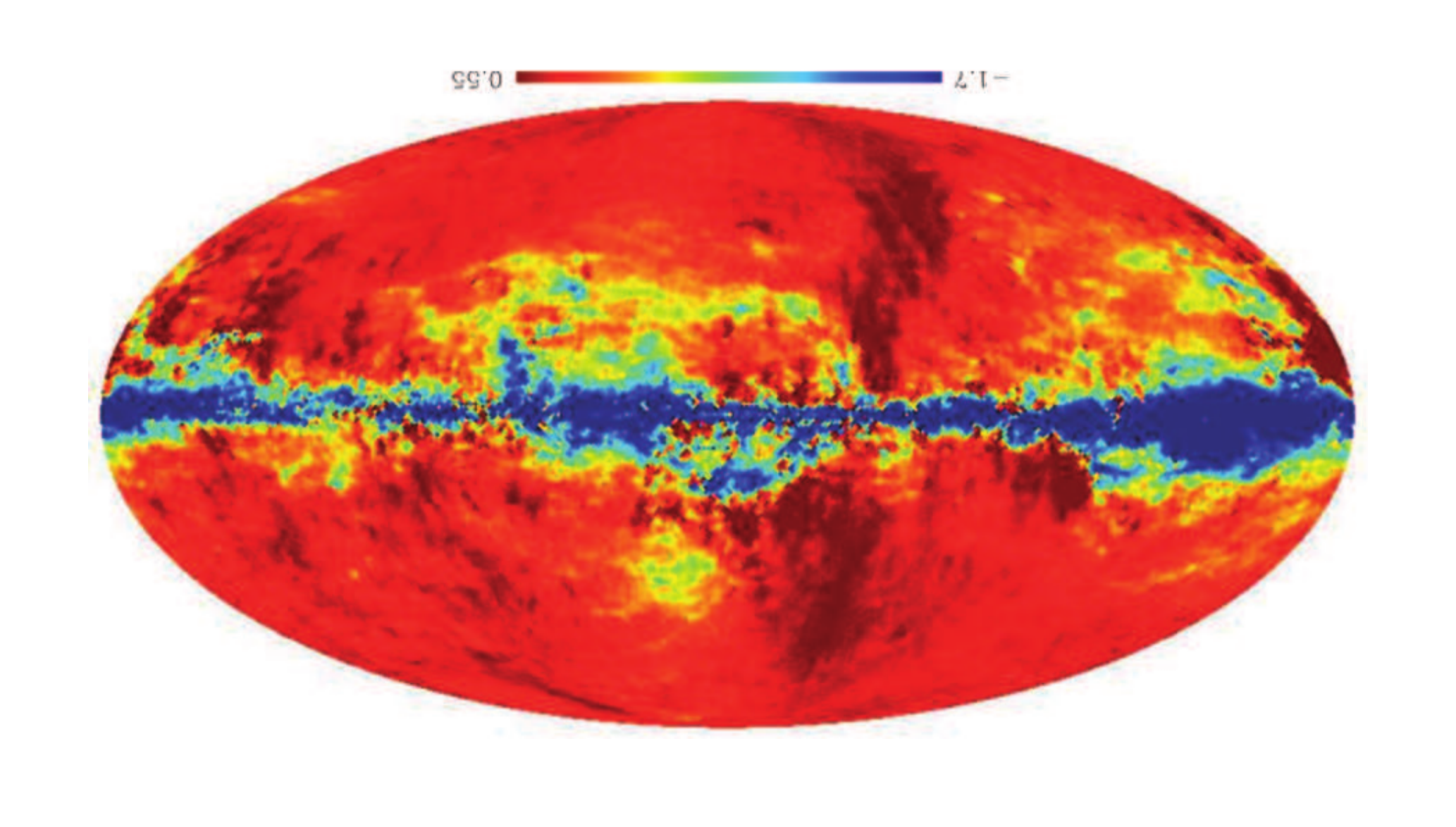}
\caption{The first of the 2 \emph{fastica} components used for the foreground model of the HM1 Q neural network. It is seen that this component both contains a Galactic component and systematic errors. The image has been converted to nside = 64. Unit: arbitrary.}
\label{fig_2}
\end{figure}


\begin{figure}[h]
\centering
\includegraphics[angle = 180, width=3.0 in]{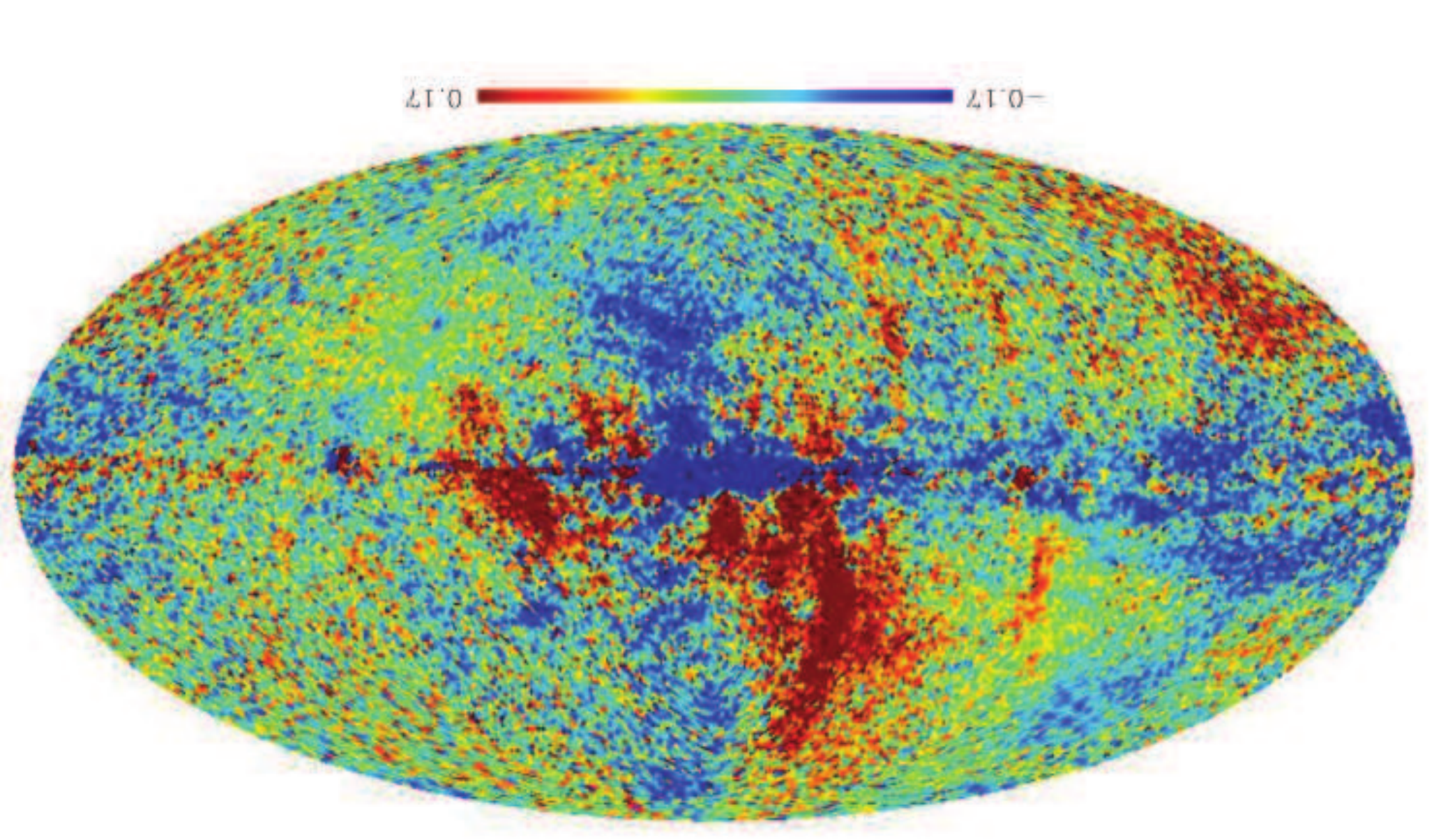}
\caption{The second of the 2 \emph{fastica} components used for the foreground model of the HM1 Q neural network. Both a galactic component and systematic errors are present in this map. The image has been converted to nside = 64. Unit: arbitrary.}
\label{fig_3}
\end{figure}

\section{The Planck polarization data}

In this investigation, only Q and U maps in the frequency range 30 GHz - 353 GHz have been analysed. As in our previous investigations, no data other than Planck data has been included in setting up the neural networks.

\subsection{The Planck observed polarization maps}

From the Planck 2015 released polarization maps, (hereafter 'Planck-set'), the following sets of maps have been used:
 \begin{itemize}
 \item{data collected from the $\textrm{1}^{st}$ and $\textrm{2}^{st}$ parts of each pointing period: HR1 and HR2}

 \item{data collected from year 1 and year 2: YR1 and YR2}

 \item{data collected in the $\textrm{1}^{st}$ and $\textrm{2}^{st}$ half of the mission (LFI years 1+3 and HFI 1) and
    in the 2. half (LFI years 2+4 and HFI 2): HM1 and HM2 }

\item{data collected from complete mission data: FULL}
\end{itemize}

The precise definition of these data sets are given in Planck Collaboration IX (2015).

\subsection{Modeling of the combined foreground spectra}
To set up the neural networks, a model of the foregrounds (in principle everything in the frequency maps, which is not CMB) is needed.
The number of frequency channels to be used in this analysis (7) puts a strong constraint on the number of parameters, which can be fitted by the neural networks. In this analysis, the interest is only to extract the CMB quantities Q and U. Therefore, a detailed physically meaningful model of the Galactic components is not required, a coherent mathematical description of the foregrounds is enough.

\begin{figure}[h]
\centering
\includegraphics[angle = 180, width=3.0 in]{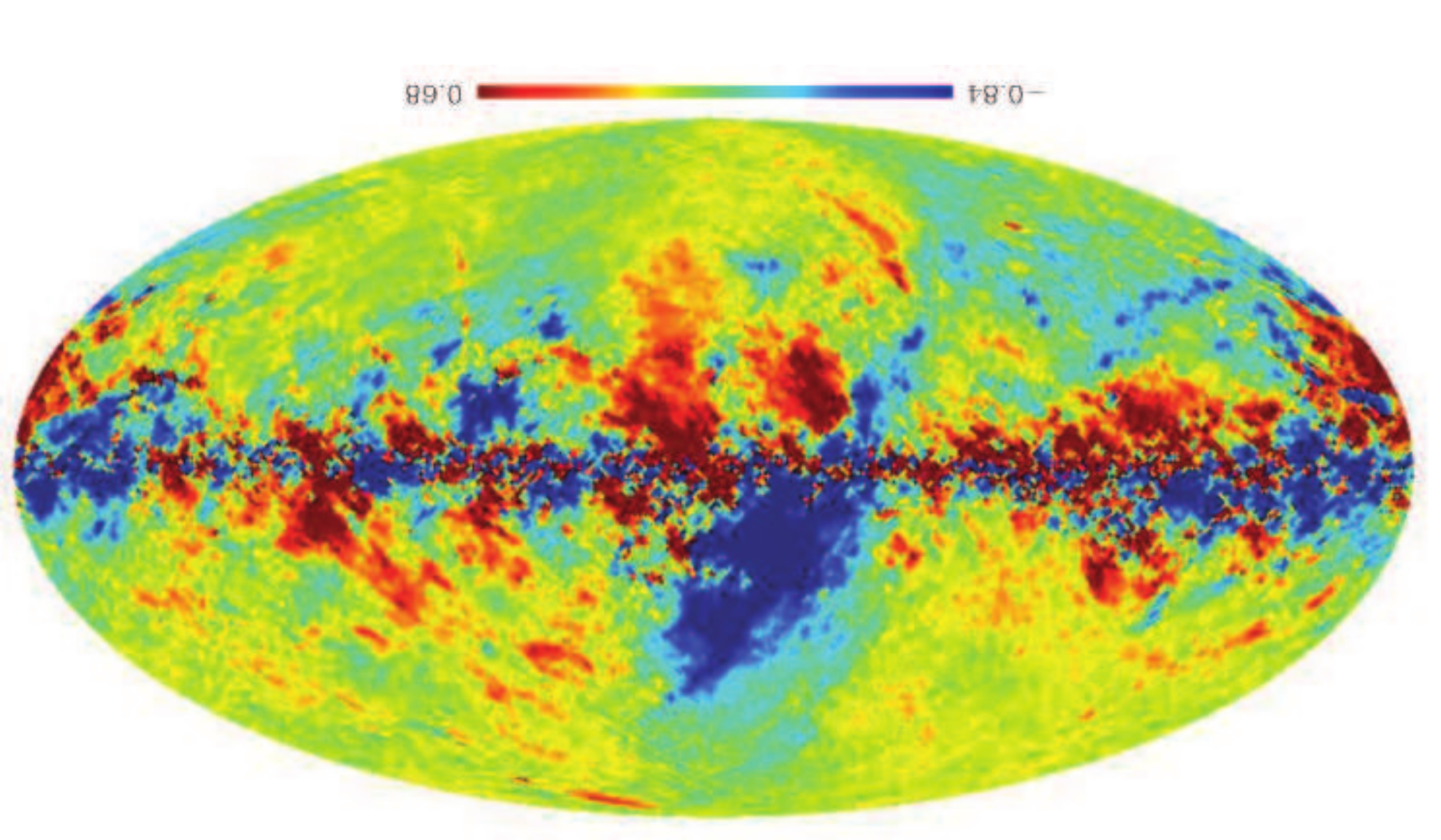}
\caption{The first of the 2 \emph{fastica} components used for the foreground model of the HM1 U neural network. Bright systematic errors and Galactic emission are present. The image has been converted to nside = 64. Unit: arbitrary.}
\label{fig_4}
\end{figure}

\begin{figure}[h]
\centering
\includegraphics[angle = 180, width=3.0 in]{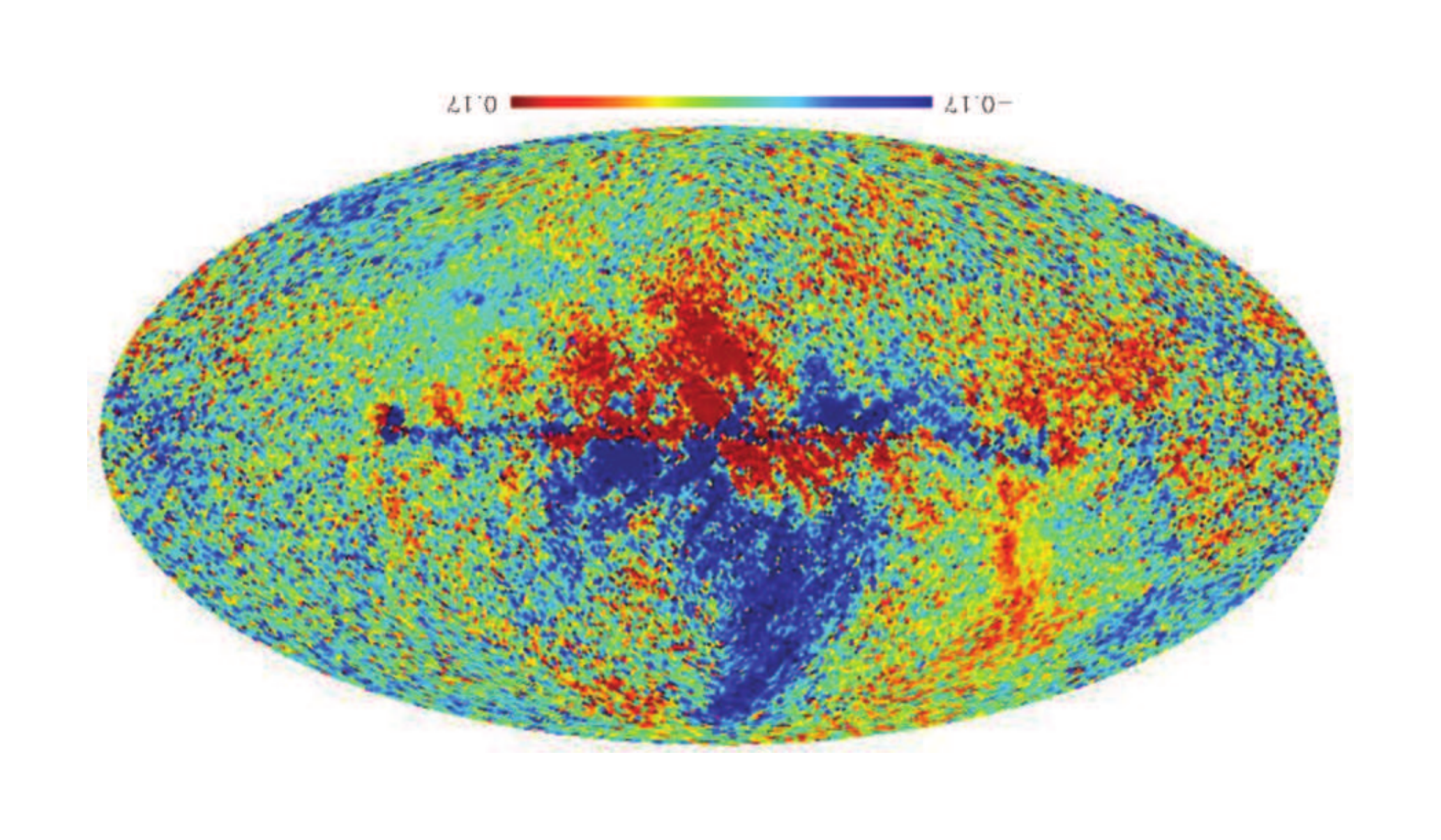}
\caption{The second of the 2 \emph{fastica} components used for the foreground model of the HM1 U neural network. Both a Galactic components and systematic errors are present in the map. The umage has been converted to nside = 64. Unit: arbitrary. }
\label{fig_5}
\end{figure}

It has turned out that the standard 'Independent Component Analysis (ICA)' method is well-suited to extract information about the non-CMB content of the maps to be analysed (an excellent description of the ICA algorithm can be found in Hyvarinen et al. 2001). The MATLAB ICA routine, \emph{fastica}, has been used to split the 7 frequency maps of each data set into 7 uncorrelated and independent components. For each component, the spectrum and the amplitude for each sky pixel is provided. Tests of the number of ICA components used to set up the foreground model have shown that  2 components contain almost all of the extended Galactic emission together with the systematic errors. From Figs \ref{fig_2}, \ref{fig_3}, \ref{fig_4} and \ref{fig_5}, the bright systematic errors in the input maps are clearly visible.

For Q and U, independent neural networks have been set up for the following data sets:
HM1, HM2, HR1, HR2, YR1, YR1, FULL.

\subsection{The Planck simulations}

The Planck FFP8 database consists of 3 components: a fiducial mission realization (comprising astrophysical foregrounds, the scalar, tensor and non-Gaussian CMB sky, and correlated instrument noise), together with separate noise and CMB realizations. The simulated maps have not been released, but details about the simulations can be found are in Planck Collaboration XII, 2015.
The FFP8 maps without 'band passband mismatch'(small differences in the effective frequency between individual detector) have been exploited. In FFP8 it has  been assumed that only synchrotron and thermal dust foregrounds are significantly polarized.

The FFP8 data used in this paper
is 1) sets of 100 independent Q and U frequency noise maps, 2) the set of foreground frequency maps and 3) the FFP8 beams for the different frequency channels.

From the CAMB software package (Lewis and Bridle, 2002, the software can be found on the webside camb.info) a set of CMB Q and U maps have been obtained, with all parameters taken with the default values, except varying the tensor to scalar ratio:  T/S = 1.0, 0.5, 0.2, 0.1, 0.0. Each CAMB CMB map has been convolved with the FFP8 beams and combined with 100 noise maps. All these maps have been run through the networks set up to analyse the Planck maps.

Furthermore, the T/S = 0.5 maps have been added to the FFP8 foreground maps and convolved with the FFP8 beams. These maps have also been combined with 100 independent FFP8 noise maps and separate networks have been established. In the following this data set will be called 'Sim-set'.


\section{The extracted Q  and U maps, high pass filtered}

It was evident for the Planck Science Team (PST) that all component separation methods investigated within the Planck Collaboration gave Q and U maps containing non-negligible remaining systematic errors at large angular scales. In order to minimize their effect on the derived power spectra, PST decided that the 'high pass' filter (hpf), shown in Table 1, should be applied to all the extracted CMB maps.


\begin{figure}[h]
\centering
\includegraphics[angle = 180, width=3.0in]{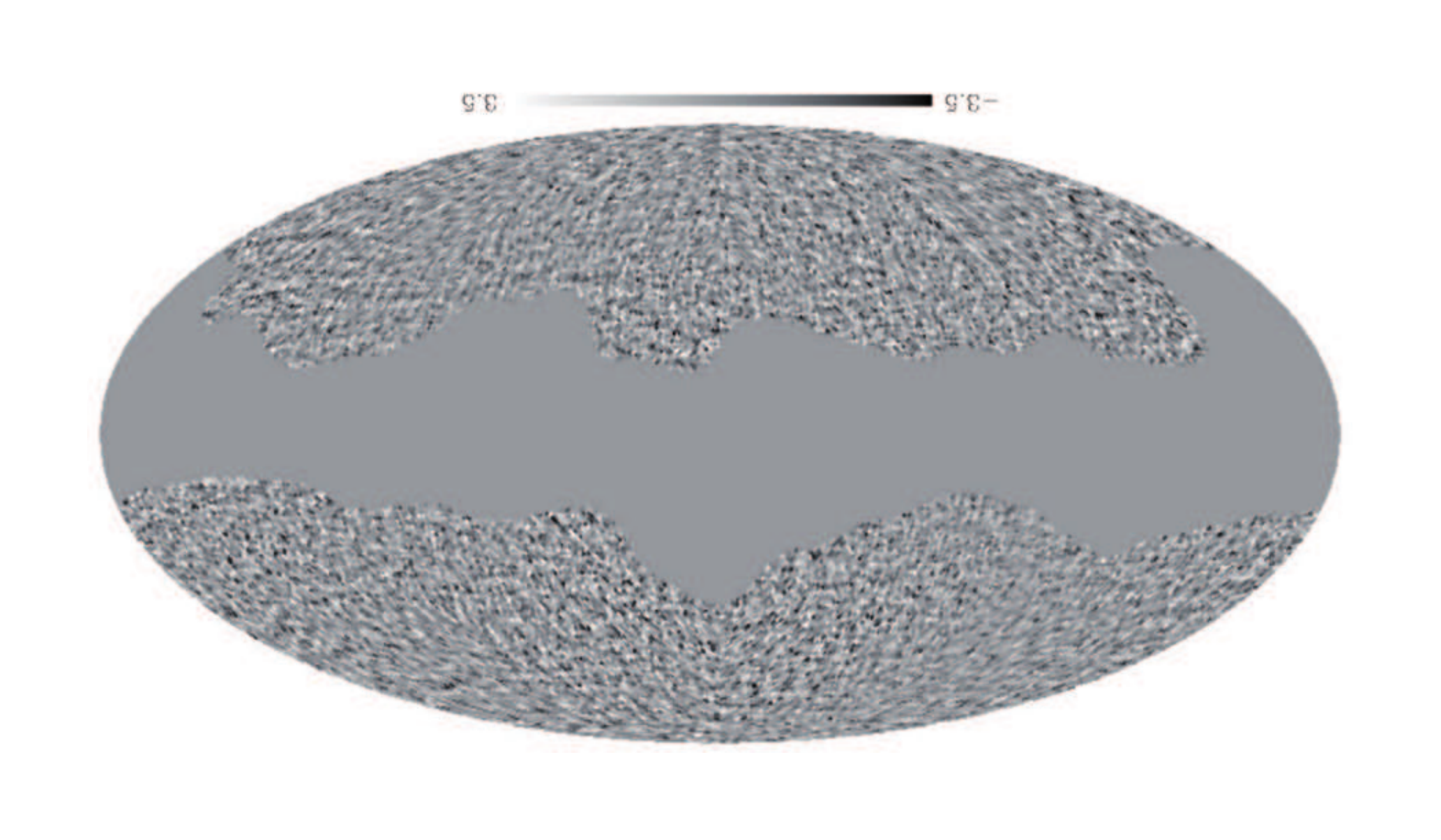}
\caption{The mean of the Q HM1 and HM2  maps extracted by dedicated neural networks, (nside = 64). Scale = $\pm$ 3.5 $\mu$K.}
\label{fig_6}
\end{figure}


\begin{figure}[h]
\centering
\includegraphics[angle = 180, width=3.0in]{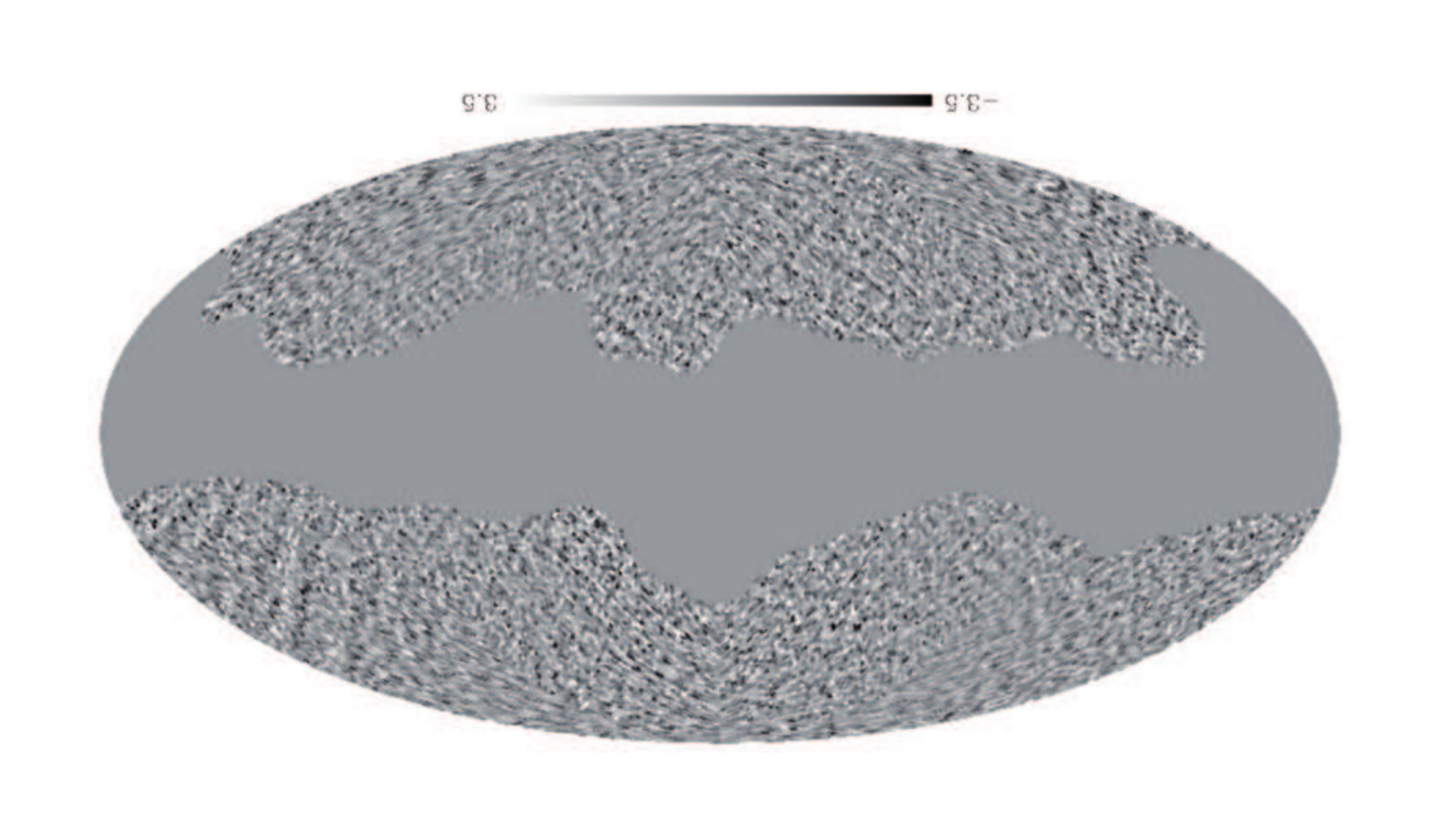}
\caption{The mean of the U HM1 and HM2 maps extracted by dedicated networks (nside = 64). Scale = $\pm$ 3.5 $\mu$K.}
\label{fig_7}
\end{figure}

\begin{table}[h]
\caption{Definition of high pass filter }
\centering
\begin{tabular}{c c}
l range & value\\
\hline
l $<$ 20 & 0.0\\
20$<$ l $<$40 & $0.5 [1~-~ cos(\pi\frac{l~-~l_{1}}{l_{2}~-~l_{1}})]$ \\
l $>$ 40 & 1.0\\
\hline
\end{tabular}
\end{table}

The mean $<$HM$>$ Q and U maps are given in Figs. \ref{fig_6} and \ref{fig_7}. It is evident that there are no visible signs of the structures seen in the \emph{fastica} components (Figs. 2 - 5). This impression is confirmed by the insignificant correlations (within the mask) between the Q and U maps and the \emph{fastica} component maps, all 4 correlations are smaller than 0.04. A few narrow black spots are found in Figs. 6 and 7, but they have no significance in the l - range relevant for the BB power spectrum.


\begin{figure}[h]
\centering
\includegraphics[angle = 180, width=3.0in]{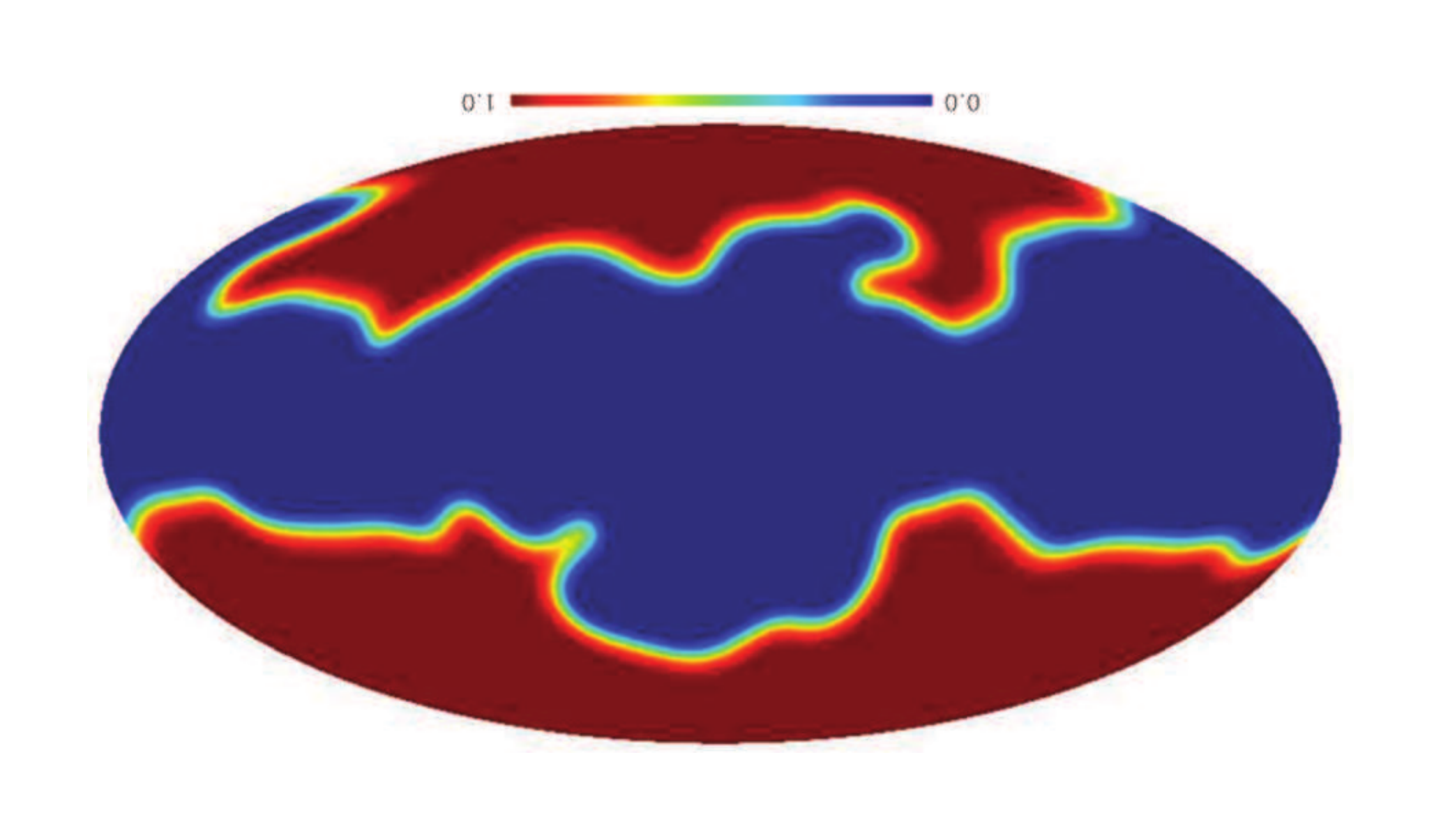}
\caption{CAMSPEC apodized mask no. 3 (Planck Collaboration XV, 2014), covering $\sim$63 \%\ of the sky.}
\label{fig_8}
\end{figure}

\section{The power spectra of the observed Planck polarization maps}

In this investigation, a mask, following the design of the CAMSPEC masks (described in Planck Collaboration XV,2014) and covering $\sim$ 63 per cent of the sky, have been applied to obtain power spectra of the extracted CMB maps. Similar masks, covering smaller part of the sky, have been investigated, but the derived power spectra showed no significant differences to the power spectrum of mask used.
Point sources are not included in this mask since they are giving a negligible contributions to the power spectra (Planck Collaboration XIII 2015).
Furthermore, introducing them would give a significant increase in the contribution of aliasing at low - l, especially in  the BB power spectra.

The Planck-set data set described in Section 2.1 has the advantage that both auto- and cross-correlation (AC and CC) power spectra of the maps can be computed. In principle, all these power spectra should give the same results, except for differences in the noise levels. So significant differences between the power spectra will give important information about remaining systematic errors and correlated noise in the maps.


All power spectra analysed in this paper have been calculated with the Fortran 90 \emph{anafast} routine, implemented in the HEALPix IDL package (Gorski et al. 2005).

\subsection{The EE power spectra}

The EE power spectra are combined in the l-intervals given in Table 2.

\begin{table}[h]
\caption{l- intervals and $\Delta$l used for the EE power spectra}
\centering
\begin{tabular}{rrr}
$l_{min}$& $l_{max}$ &  $\Delta$l\\
\hline
50 & 1100 & 10\\
1100& 1800 & 50\\
1800& 3000 & 100\\
\hline
\end{tabular}
\end{table}


\begin{figure}[h]
\centering
\includegraphics[width=3.0 in]{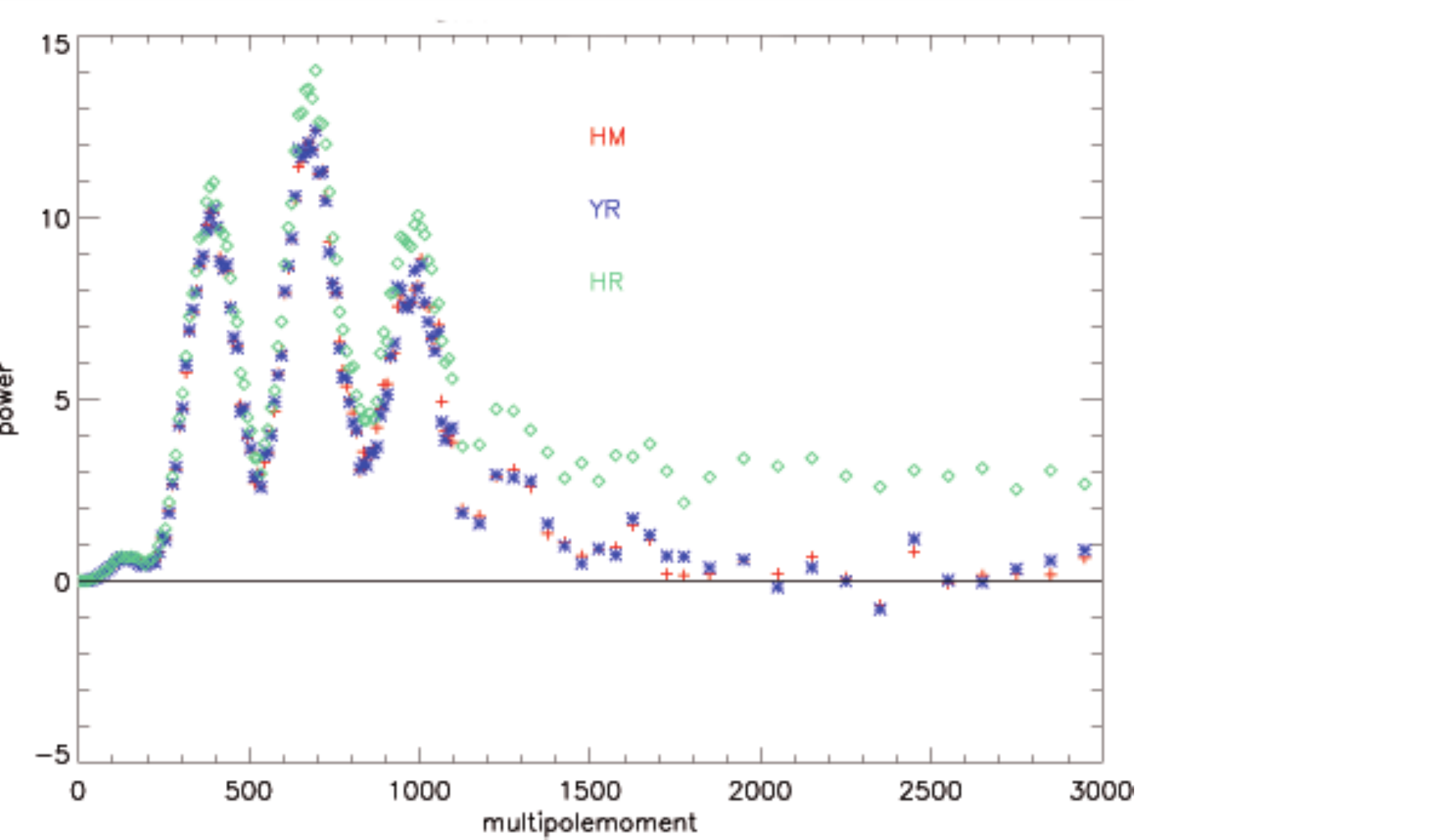}
\caption{The CC EE power spectra for the HM, HR and YR data sets. It is seen that HR spectra contains significant amounts of correlated noise.}
\label{fig_9}
\end{figure}

In Fig. \ref{fig_9} it is seen that the CC EE HR power spectrum has a much higher level at l $>$ 1100, implying a significant amount of correlated noise in the HR1 and HR2 data sets. Due to this correlated noise, the HR spectrum is not further discussed. As seen in Sect. 3.1 the HM data sets cover a larger period than the YR data sets, implying a somewhat better noise level. Consequently, the HM data sets are chosen as the baseline in the following.

Before it is possible to get useful power spectra out of the AC power spectra, an accurate noise power spectrum needs to be subtracted. This spectrum has been obtained by averaging the two difference power spectra (HM 1$-$2 and YR 1$-$2). Since this combined spectrum contains no features, it has been median filtered (nmed = 25). With this construction of the noise spectrum, it is assured that the noise level of the derived AC spectra is not increased significantly.

The AC EE FULL noise subtracted spectrum correlates well ($~$ 0.99) with the CC EE HM spectrum and will not be discussed in detail.

\subsection{The BB power spectra}

All BB power spectra have been combined in multipole intervals of $\Delta$l = 25.

\begin{figure}[h]
\centering
\includegraphics[width=3.0in]{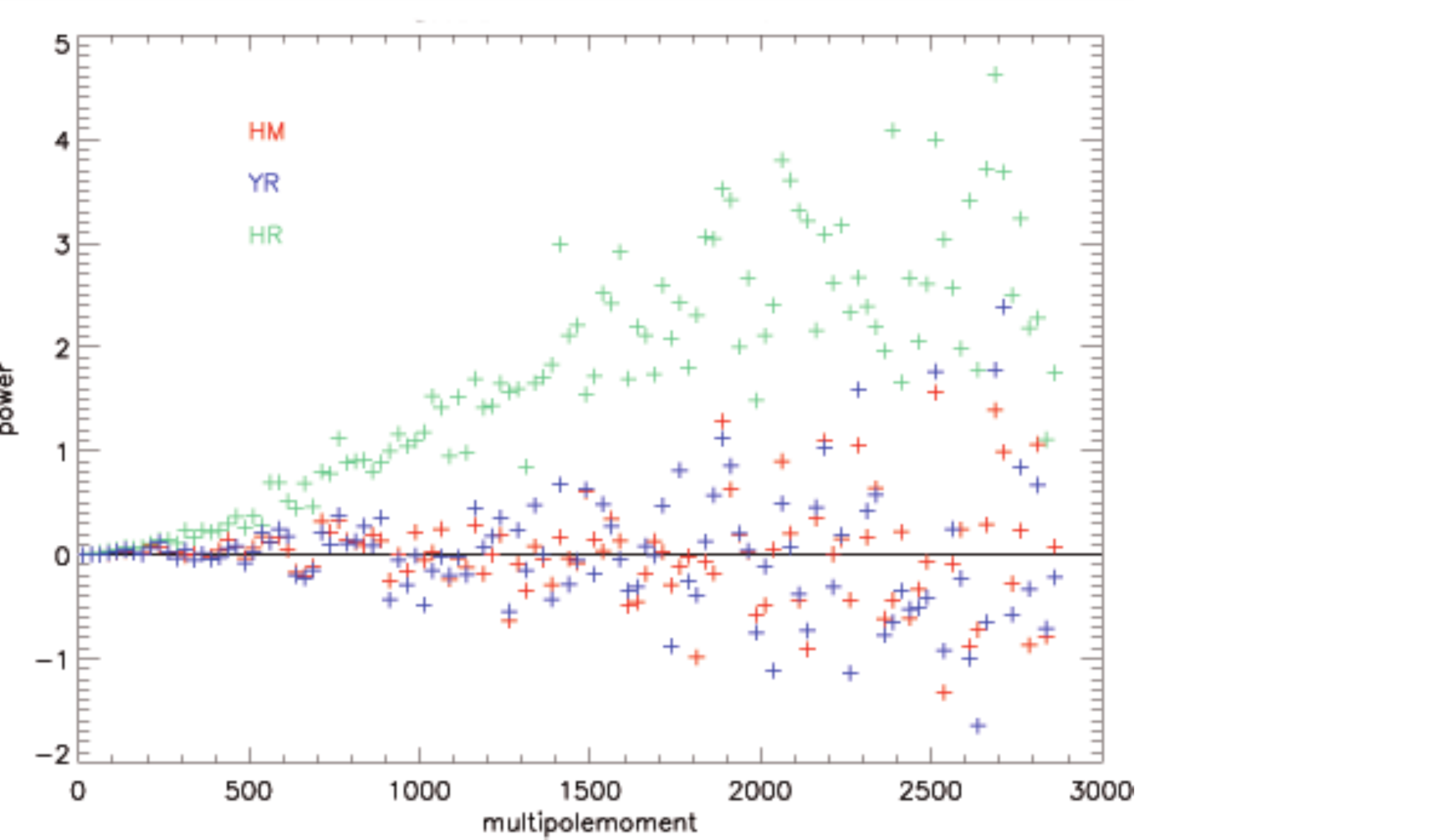}
\caption{The CC HM, YR and HR BB power spectrum.
It is seen that the YR spectrum fits well to HM, while the HR shows similar trends as the CC EE HR spectrum. }
\label{fig_10}
\end{figure}

Fig.\ref{fig_10} shows the CC BB HM, YR  and HR power spectra. It is seen that CC HR spectrum shows a similar trend as the CC EE HR spectrum (see Fig. \ref{fig_9}). Therefore,  the CC BB HR power spectrum will not be further discussed in this paper.


As for the mean EE noise power spectra, the mean BB AC noise spectrum has been established by combining the power spectra of the difference maps HM 1-2 and YR 1-2.
There is no significant difference between the CC BB HM and AC HM spectra in the range 50 $\leq ~l~ \leq$ 400. Therefore, only the CC BB HM spectrum will be discussed in the following.

\begin{figure}[h]
\centering
\includegraphics[width=3.0 in]{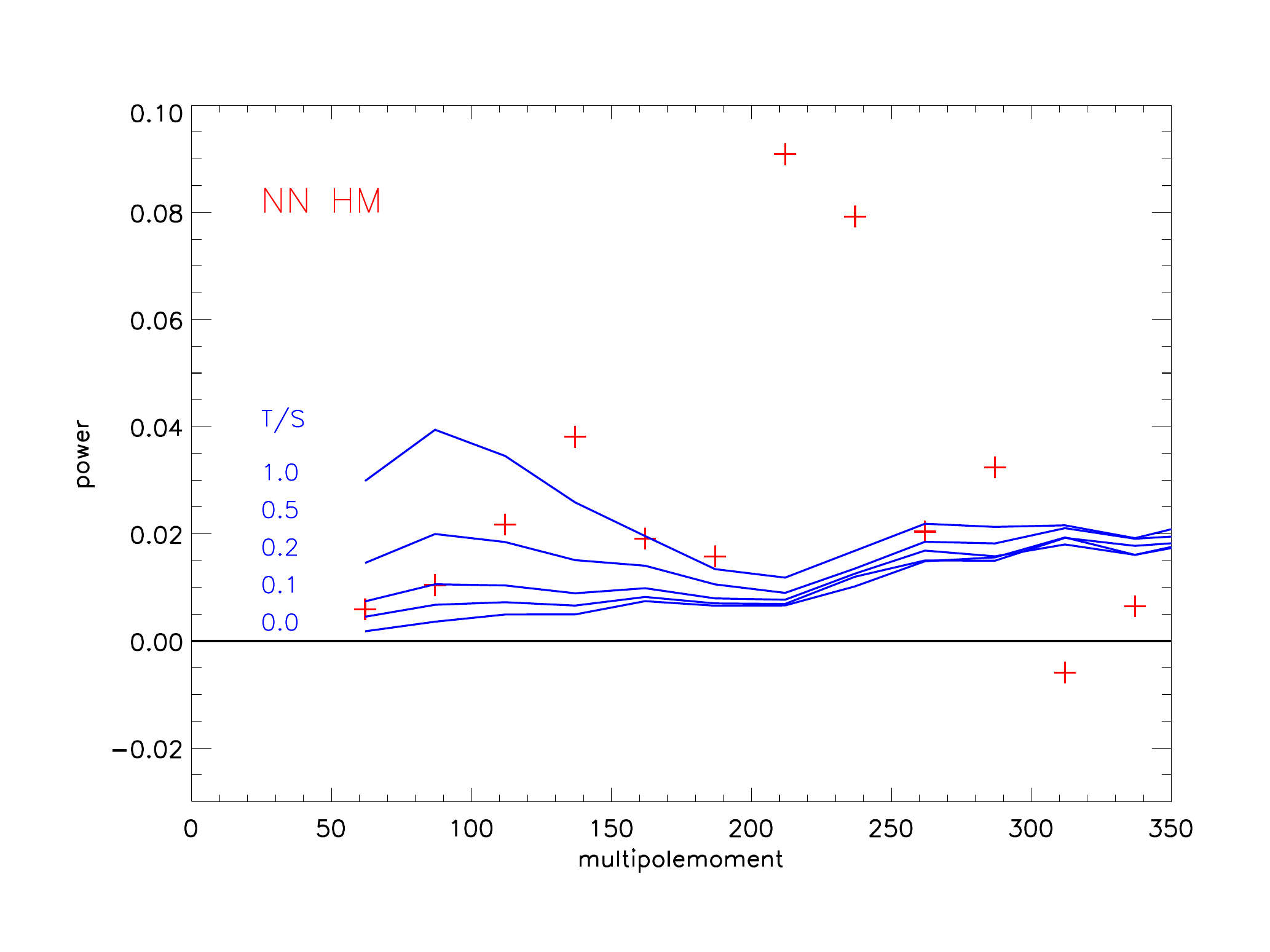}
\caption{The HM BB power spectrum compared with the derived  'Sim-set' BB spectra, with no corrections applied. It is seen that the 'Sim-set' spectra form, as expected, a group with the T/S ratio as the main parameter.}
\label{fig_11}
\end{figure}

Fig. \ref{fig_11} shows the 5 raw (no corrections applied) CC BB CAMB spectra, together with the raw CC HM BB spectrum. As expected, it is seen that the CAMB spectra are basically a group with one parameter, namely T/S. These spectra are calibrated (Sect. 6) by multiplying corrections for the mask and for the effective window function, implying that the relative strengths of the spectra are preserved. It is also evident that the HM BB spectrum will be difficult to fit within this group of spectra.


The CC EB HM power spectrum shows,  as expected, no significant deviations from zero.

\section{The calibration of the power spectra}

The calibration is performed in the following steps:
\begin{itemize}
\item{corrections for the differences in the power spectra due to inclusion of the foregrounds in the simulated maps}
\item{corrections for the size of the sky mask}
\item{corrections for the effective window function of the neural networks}
\end{itemize}

\subsection{The EE power spectra}
\begin{figure}[h]
\centering
\includegraphics[width=3.0in]{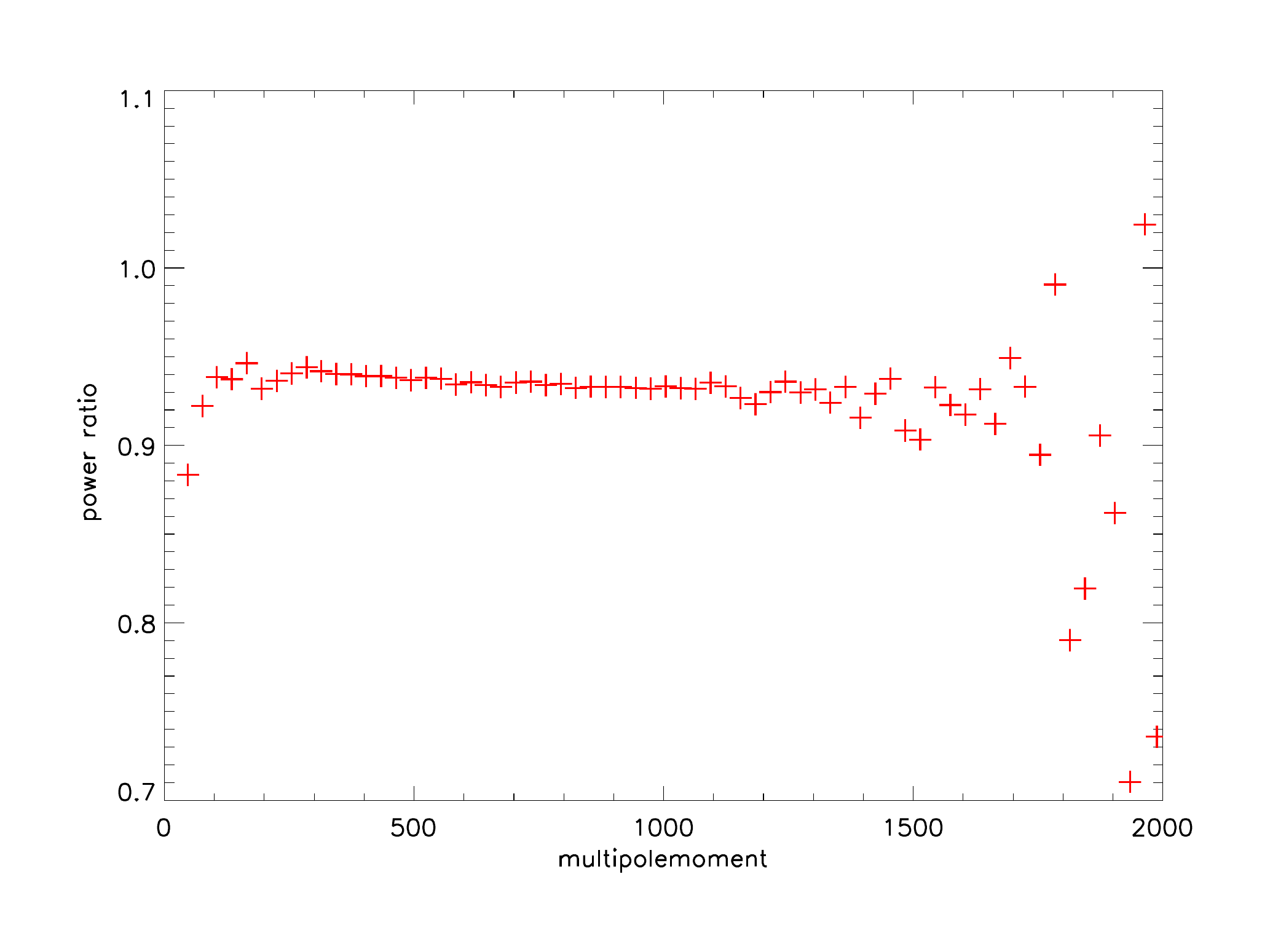}
\caption{The ratio between the derived 'Sim-set' EE T/S = 0.50 'CMB + noise' spectrum and the same with the foreground included in the data set, derived with an independent neural network.}
\label{fig_12}
\end{figure}

Although a lot of efforts has gone into defining the 'Sim-set' as closely as possible to resemble the real sky,
small deviations are probably still present. It is only possible to test the performance of the 'Planck-set' networks by using the 'Sim-set CMB only' maps, while
the 'Sim-set cmb + foregrounds' networks can be tested with both the 'CMB only' and 'CMB + foregrounds' maps. From these power spectra, information about the influence of the foregrounds can be estimated. Figs. \ref{fig_12} shows the ratio between the power spectra for 'CMB only' and for 'CMB + foregrounds'.

\begin{figure}[h]
\centering
\includegraphics[width=3.0in]{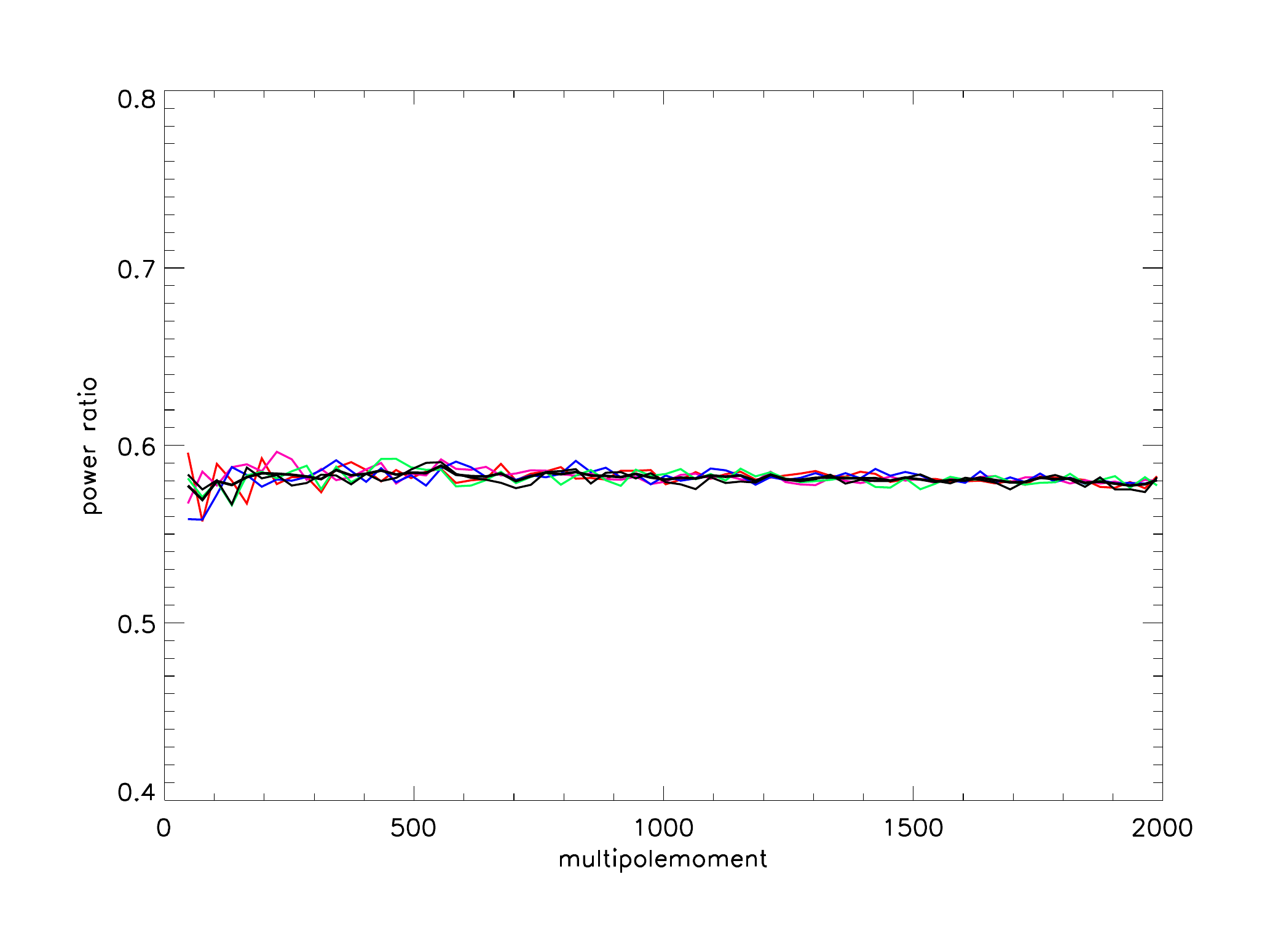}
\caption{The figure shows the EE aliasing corrections (the ratio between the power spectra of intrinsic CMB maps with and without the mask applied) obtained for each of 5 CAMB data sets, shown with different colors. The black curve is the mean of the 5 corrections.}
\label{fig_13}
\end{figure}

The corrections for the size and shape of the sky mask (Fig. \ref{fig_8}) have been calculated as the ratio between of the power spectra of the  intrinsic CMB map with and without applying the mask.
Fig.\ref{fig_13} shows these corrections. The mean of these corrections have been applied (black curve in Fig. 13).


\begin{figure}[h]
\centering
\includegraphics[width=3.0in]{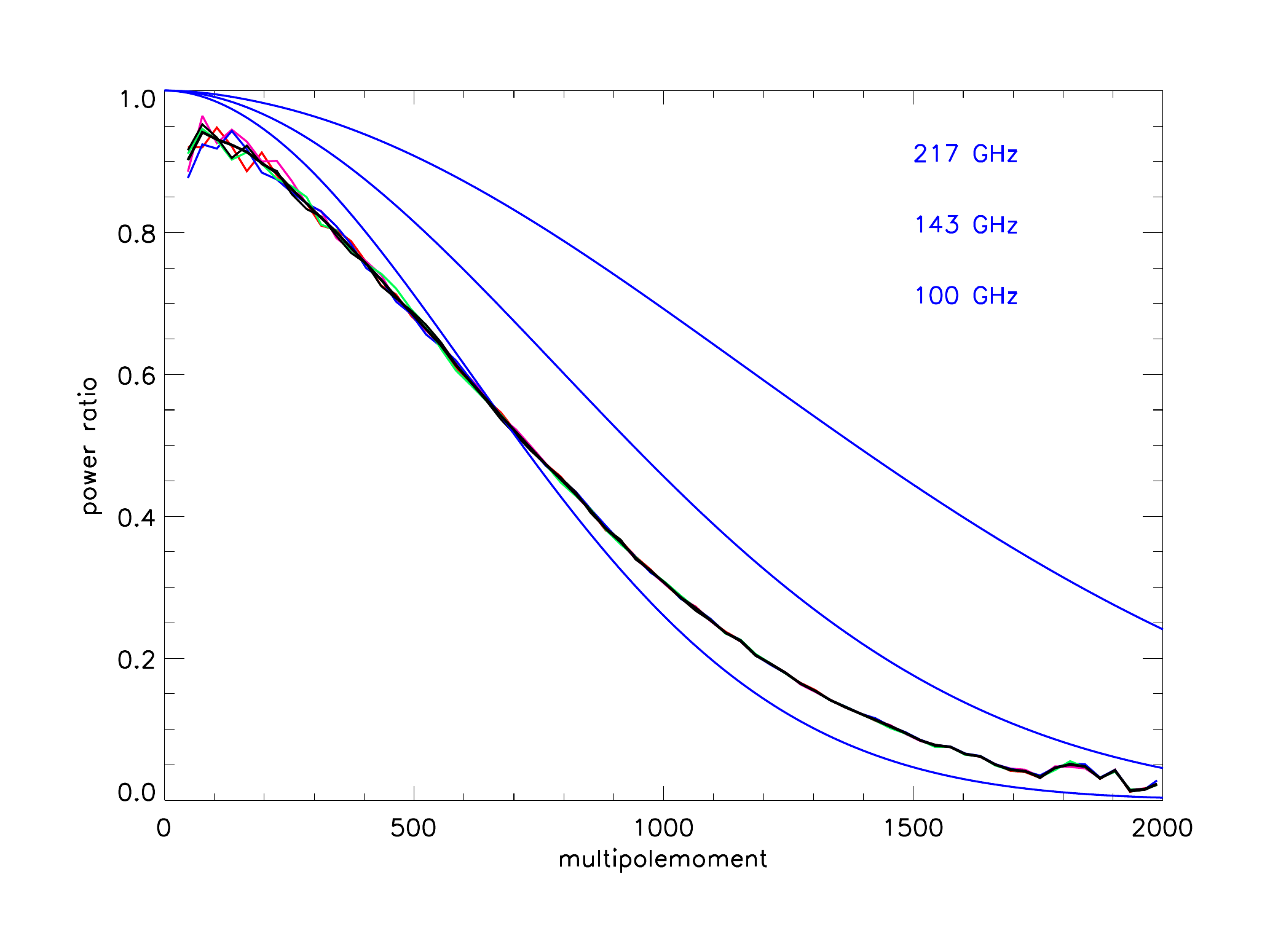}
\caption{The figure shows the EE effective window functions (the mask - corrected CAMB spectra divided with the power spectrum of the intrinsic CMB map) obtained for each of the 5 CAMB data sets, shown with different colors. The black curve shows the mean of the 5 CAMB functions. The blue curves show the 100 GHz, 143 GHz and 217 Gz window functions applied in the 'Planck-set'}
\label{fig_14}
\end{figure}

As emphasized above, the frequency maps run through the neural networks are not corrected for the differences in the instrumental beams. To determine the effective window functions of the
networks the 'Sim-set' of 100 'CMB only' data sets are run through the 'Planck-set' networks. The mean power spectra are corrected for the mask and then compared with the 'Sim-set' intrinsic CMB spectra. The effective EE window functions are shown in Figs.\ref{fig_14} and the mean is used.

\subsection{The BB power spectra}
\begin{figure}[h]
\centering
\includegraphics[width=3.0 in]{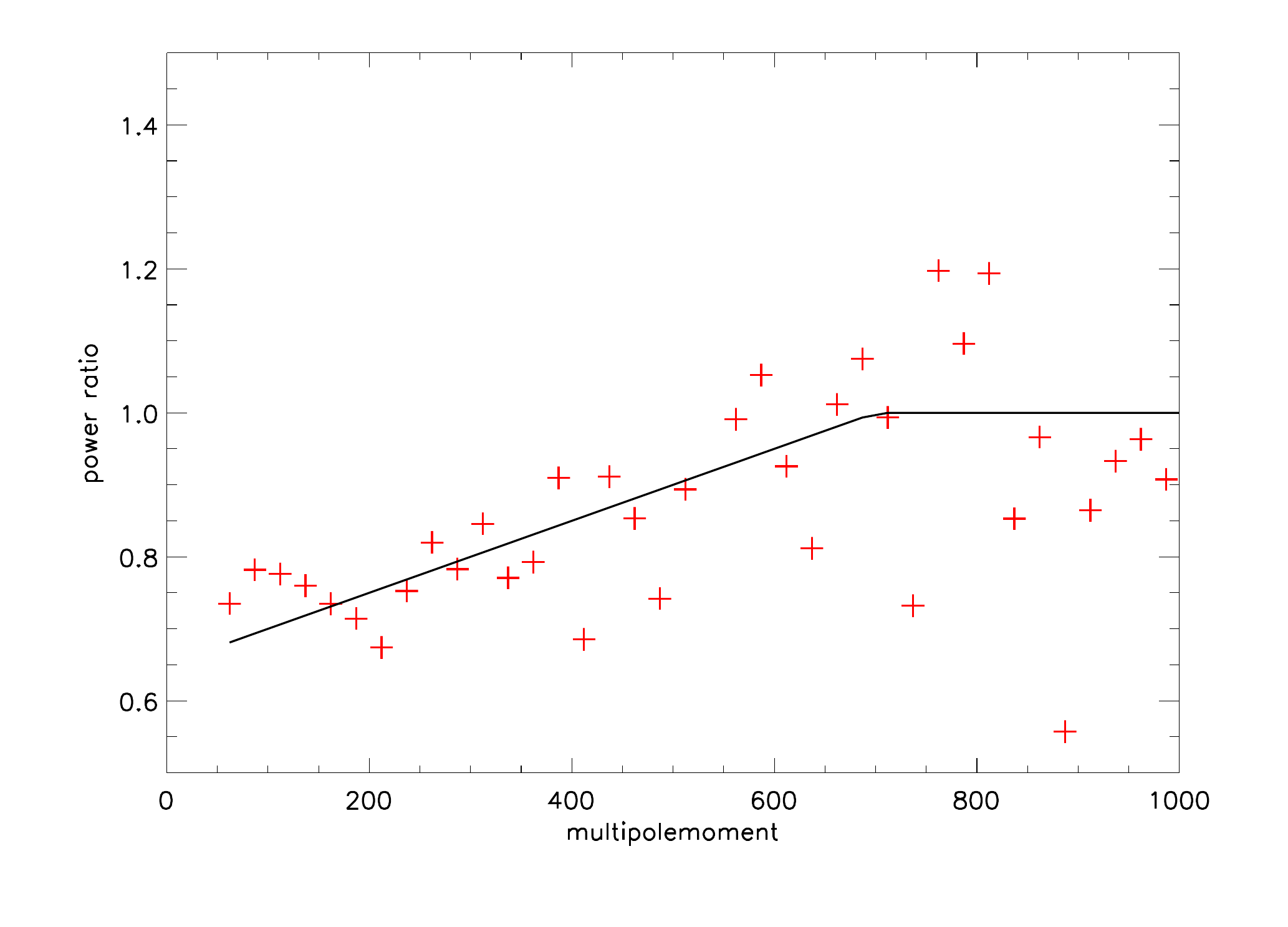}
\caption{The correction of the BB power spectra for the difference of the derived power of the  'Sim-set' (T/S=0.5) with and without including the foregrounds. It is evident that the bright feature found in the HM BB power spectrum for 200 $\leq$ l $\leq$ 250 (see Fig. 17) is not present in the 'Sim-set CMB + foreground' spectrum.}
\label{fig_15}
\end{figure}

The BB power spectra are calibrated using 'Sim-set' in the same scheme as the EE power spectra. As seen in Fig. \ref{fig_11}, the signal of the T/S = 0.0 and 0.1 is low, implying a larger uncertainty than for the others. Therefore,  the corrections for the BB power spectra are averaged over the 3 highest T/S spectra. Fig. \ref{fig_15} shows the ratio between the 'CMB + noise' and the 'CMB + foregrounds' power spectra for the T/S = 0.5 data sets.
The  strong feature found in the HM BB power spectrum for 200 $<$ l $<$ 250 (see Fig. 17) is clearly not present in the 'CMB + foreground' spectrum, implying that these corrections are not responsible for the bright feature at l $\sim$ 225 in the HM BB power spectrum.


\begin{figure}[h]
\centering
\includegraphics[width=3.0in]{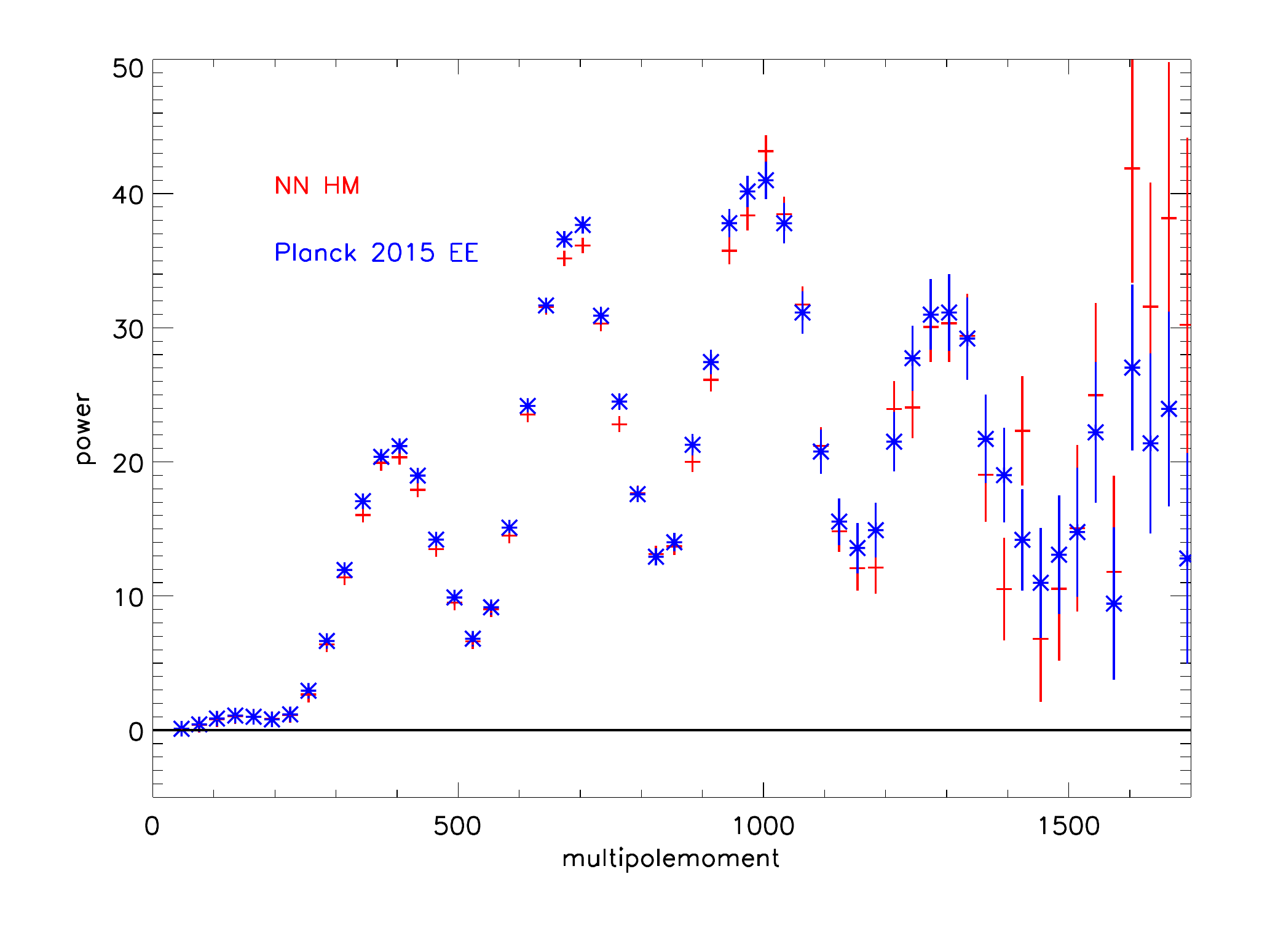}
\caption{The fully calibrated EE HM cross - power spectrum  compared with the Planck 2015 EE spectrum. It is seen that the two spectra fit together up to l $\sim$ 1600. The error bars are derived from the rms around the mean power spectra derived from the CAMB data sets.}
\label{fig_16}
\end{figure}

\section{The calibrated EE and BB power spectra}

 Estimates of the noise levels in the calibrated power spectra have been found from the extracted 'Sim-set' power spectra as the rms around the mean of  the EE and BB spectra.
 Since this paper is mainly dealing with the signal-to-noise of the calibrated power spectra (not the theoretical interpretation), the cosmic variance is not included in deriving the error bars.


\begin{figure}[h]
\centering
\includegraphics[width=3.0 in]{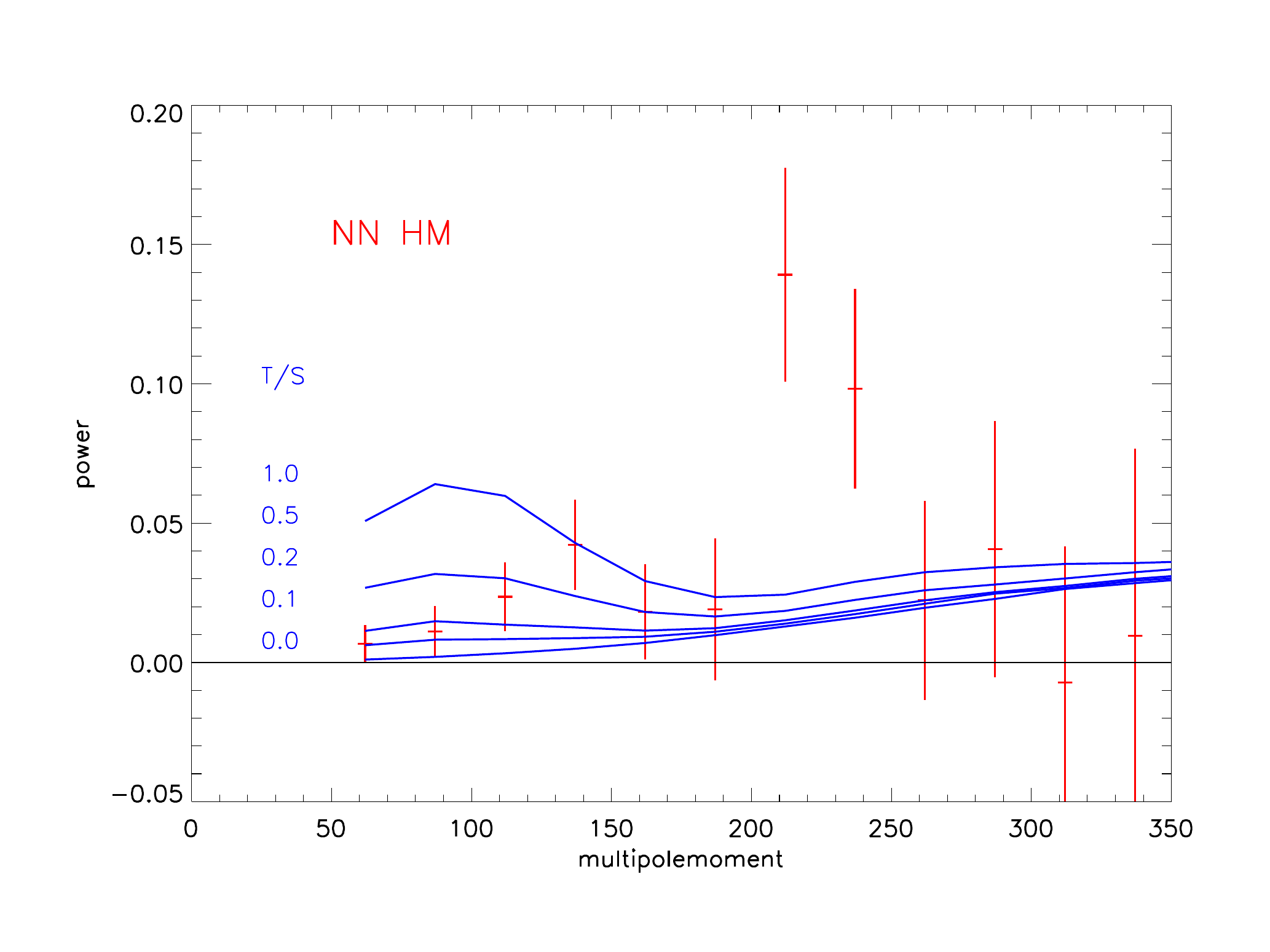}
\caption{The fully calibrated HM BB power spectrum, The error bars are derived from the rms around the mean power spectra derived from the CAMB data sets. The CAMB intrinsic BB spectra are also shown. It is evident that this group of models fit poorly to the HM BB spectrum. The bright feature around l $\sim$ 225 has a S/N of 4.5 .}
\label{fig_17}
\end{figure}

The fully calibrated EE HM power spectrum  and the Planck 2015 EE spectrum are shown in Fig. \ref{fig_16}. The 2 spectra are in excellent agreement, for 100 $\leq$ l $\leq$ 1400 the correlation is 0.99 .

The fully calibrated CC HM BB power spectrum is shown in Fig. \ref{fig_17} together with the intrinsic BB power spectra of the 'CAMB data sets'. It is evident that HM BB spectrum fits badly with the CAMB spectra. The bright feature for 200 $\leq$ l  $\leq$ 250 has a S/N $\simeq$ 4.5.

\begin{figure}[h]
\centering
\includegraphics[width=3.0 in]{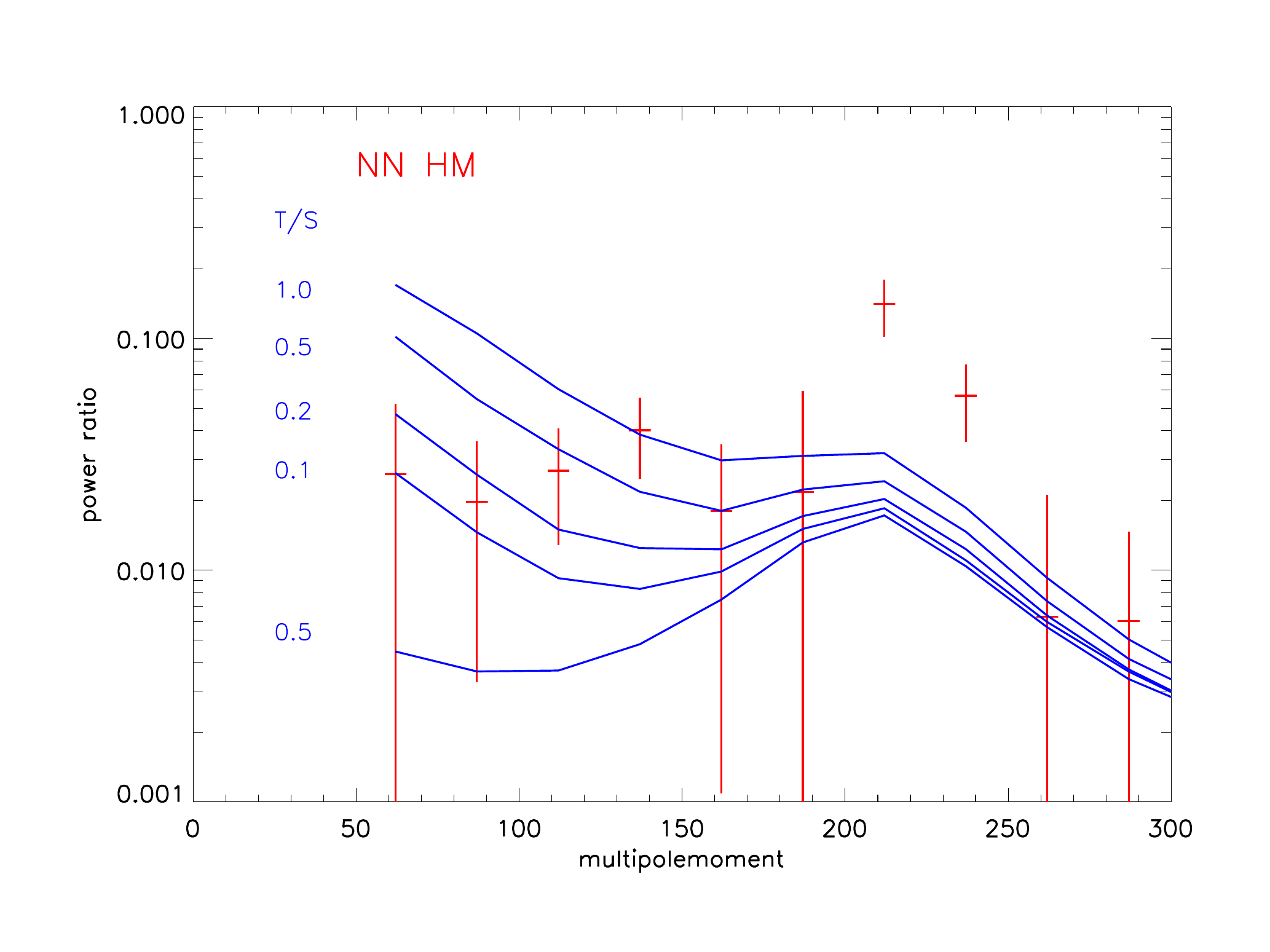}
\caption{The ratio of the HM BB  and EE power spectra compared with the ratios of the CAMB intrinsic spectra. The error bars are derived from the rms around the mean power spectra derived from the CAMB data sets.}
\label{fig_18}
\end{figure}

The ratio of the HM BB  and EE power spectra is shown in Fig. \ref{fig_18}. Similar ratios are also plotted for the intrinsic CAMB spectra. Although the HM ratios are within the same order of magnitude as for these theoretical models , they fit poorly together.

It is striking that the NN HM BB power spectrum are significantly brighter than previous results.

 Ade et al. (2015) determines a scaled $\lambda$CDM lensing spectrum (scaling factor of 1.12$\pm$0.18) and an upper limit on r $\leq$ 0.13 (95 \%\ confidence) for a BB-modes spectrum. The BICEP2 sensitivity in this small area of the sky is more than a magnitude higher than what Planck has achieved in the same area (see e.g.  Ade et al. Fig. 13), implying this is basically a BICEP2 result, where Planck has provided a statistical estimate of the dust emission.

The analysis in Planck Collaboration XIII (2015) is based on power spectra derived from cross-correlation spectra of individual frequency maps (100 GHz, 143 GHz and 217 GHz) for individual detector combinations, within sky masks covering from 31 to 49 \%\ of the sky. As seen in Figs.\ref{fig_2} to \ref{fig_5}, there are significant remaining systematics errors in the frequency maps even for these 'small' masks.

The advantages of the neural network method is outlined above: a small number of weights is necessary to derive the desired CMB quantities, only the Planck observed and simulated data used, small number of parameters used to calculate the Galactic emission, search for features with a spectral behaviour as CMB, and all frequency channels exploited.

As demonstrated in our previous papers, the neural networks themselves are not introducing non - physical features in the extracted maps.
Furthermore, in Fig. \ref{fig_11} it is seen that the method has not introduced a bright l $\sim$ 225 feature in the 'Sim-set CMB + noise' BB power spectra. From Fig. \ref{fig_15} it can be seen that the same is the case for 'Sim-set CMB + foregounds' power spectra. This figure is also showing this feature is not present in the 'Sim-set' foreground models.

All together, this feature must originate in a faint component on angular scales of $\sim$1 deg and a frequency spectrum similar to CMB. With these characteristics, it will be a basic problem for all component separation methods to distinguish it from the real CMB.

At this stage it is difficult completely rule out a non - CMB origin of this feature. Further analysis will have to wait until the final Planck data are released.

\section{Conclusions}
The feasibility of neural networks for extracting CMB from mm/submm polarization observations has been extended to the latest available set of data from the Planck satellite.

Although a lot of efforts have been performed within the Planck Collaboration in order to remove correlated noise and systematic errors from the 'Planck-set', a significant amount of  systematic errors is still present. It has been demonstrated that the level of systematics in the Q and U maps has been reduced to a very low level by the neural networks in combination with the 'high pass filter'.

A series of data sets including CAMB spectrum with different tensor to scalar ratios (T/S = 1.0, 0.5, 0.2, 0.1, 0.0) and a lensing spectrum ('Sim-set') has been used to calibrated the power spectra of the derived maps by the neural networks.

The final EE power spectrum fits very well to the Planck 2015 spectrum for l $\leq$ 1600.

The final BB power spectrum, derived by the neural networks, shows detection with a signal to noise ratio of 4.5 for $200\leq~l~\leq 250$. It is clear that this BB spectrum fits poorly with these  CAMB  spectra. This feature is characterised  as  a faint component on the Sky, angular scales of $\sim$1 deg, and a frequency spectrum similar to CMB. For such a component, it will be a basic problem for all component separation methods to distinguish it from the real CMB.

A firm conclusion on this feature has to wait  until the final Planck maps have been released.

\acknowledgements
The author acknowledges that this work would not have been possible without the dedicated efforts by a lot of strongly committed scientists and engineers within the Planck  Collaboration.  Dr. C. A. Oxborrow is acknowledged for most valuable comments to the manuscript.

\end{document}